# Nature of magnetism and transport in $B_xC_yN_z$ thin films: The intriguing role of nitrogen defects in the electronic structure

Rajesh Mandal[1], Rohit Babar[1,], Malvika Tripathi,[3] Shouvik Datta,[1,2] R. J. Choudhary,[3*] Mukul Kabir[1,2*] and Satishchandra Ogale[1,2*]

[1] Department of Physics, Indian Institute of Science Education and Research, Pune, India-411008

[2] Centre for Energy Science, Indian Institute of Science Education and Research, Pune, India-411008

[3] UGC-DAE Consortium for Scientific Research, Indore Centre, University Campus, Khandwa Road, Indore 452017, India

## Abstract

Boron carbonitride (BxCyNz) represents an interesting family of materials containing all light elements and two dimensional graphene like hybrid layers. Although rich literature exists on this peculiar material in chemically processed form, there are relatively fewer reports on device application-worthy thin film form. This form also offers a natural platform for basic studies on important aspects such as magnetism and transport, which have attracted attention in the context of ferromagnetism in doped and defect carbon systems. Thus, in this work we have grown thin films of this compound using Pulsed Laser Deposition (PLD) method, and investigated their magnetic and transport properties in their entirety along with the detailed electronic structure using various spectroscopic techniques like X-ray photoelectron spectroscopy (XPS), Valence Band Spectroscopy (VBS) and X-ray absorption near edge spectroscopy (XANES). In depth analysis of the typical role of dopants and defects, especially the prevalent nitrogen defects, is elucidated using Density Functional

Theory (DFT) calculations to understand the experimental observations. A dramatic crossover in the transport mechanism of charge carriers is observed in this system with the change in doping level of specific nitrogen defects. A robust and high saturation magnetization is achieved in BCN films which is higher by almost hundred times as compared to that in similarly grown undoped carbon film. An anomalous transition and increase in coercive field is also observed at low temperature. A careful analysis brings out the intriguing role of specific nitrogen defects in defining the peculiar nature and concentration dependence of the physical properties of this system.

## I. INTRODUCTION

Possibility of inducing or realizing magnetism and tunable electronic structure in carbon systems has always attracted significant attention of experimentalists and theorists alike for quite some time, because of the uniqueness and diversity of carbon forms available in nature from a highly anisotropically conducting graphite (sp2 bonding) to highly insulating diamond ($sp^3$ bonding), with a variety of intermediate allotropes spanning all dimensionalities (fullerenes, CNTs, graphene, turbostratic carbon etc.); thanks to the magical display of the admixtures of $sp^2$ and $sp^3$ hybridized C-C bonds. Most efforts in this field have been directed towards inducing a magnetic moment in a specific functional carbon form by synthetic incorporation of one or more dopant or defect types in the carbon system, and exploring the possibility of connecting the moments by exchange interactions either via conduction electrons or direct exchange. Dilute dopant concentration induced ferromagnetism in otherwise non-magnetic systems (e.g. diluted magnetic semiconductors) has separately been a subject of great scientific interest for quite some time[1]. Other forms of conducting carbons such as graphene ribbons have provided intriguing possibilities of magnetism induced by edge state manipulation[2]. It is important to emphasize that in many high surface area forms of functional carbons, it is also possible to functionalize the surfaces by molecular attachments, providing further options in this respect[3].Unfortunately, the possibility of a variety of options for the choice carbon forms and manipulating them for studies of magnetism has also been detrimental to the development of a cohesive and clear picture about the mechanisms in this context. The literature has therefore witnessed reports that are very interesting in their own right but conflicting and confusing in different ways. Moreover the materials platforms that have been examined are also equally diverse (ranging from powders to films), making it impossible to compare experimental results. Also, the connection between thorough atomistic characterization of the microstates of the given system, transport properties and magnetism is not always sought concurrently both experimentally and theoretically. This scenario clearly invites research that could be performed on robust materials, methods, and platforms that are synthetically highly controlled and have a connection to possible technological utilization, if a robust ferromagnetism can be realized therein. The present work is a modest step along this direction.

Herein we have explored the possibility of realizing ferromagnetism and its connection with the electronic structure in pulsed laser deposited (PLD) nitrogeneted turbostratic carbon films without and BN doping. The choice of nitrogen doping is guided by the rich literature already available on this subject, while the choice of BN doping is attempted to establish the possible connection to the novel 2D material $B_X C_Y N_Z$(BCN) that has attracted significant attention lately. We have however restricted to low boron doping regime to restrict the analysis to doping scenario than an altogether new material, as far as possible. Importantly, we have used a variety of key atomistic characterizations along with transport and magnetism measurements, as well as first-principles theoretical studies to bring out a fairly detailed picture pertaining to the evolution of relationship among the structural, electronic and magnetic properties in this specific and limited domain of vapor phase (PLD) deposited turbostratic carbon films. This form has graphitic layers of short lateral areal dimension that are not as neatly stacked as in graphene, as discussed later. An important aspect is that this form can be controlled in the film form by controlling the growth methods and conditions, and hence highly reproducible results can be obtained. Moreover thin film form is akin to technological utilization including novel spintronic device in stand-alone form or via coupling with other materials that can also be grown by vapor phase methods in hetero-structure configurations[4].

Before proceeding to present and discuss our results it is useful to highlight some of the interesting theoretical and experimental reports in the literature which have addressed this problem of (ferro) magnetism as well as transport properties in carbon without and with N and BN doping in different ways, with a view to seek valuable insights for the analyses of our data; although as stated earlier, it is really hard to compare and contrast results very strictly in this field in view of the diversity of forms that have been examined.

The possibility of inducing magnetic moment via defects and vacancies in graphene and graphitic carbon system has been explored theoretically by many researchers. Faccio et al[5] showed the evolution of magnetic moments by inducing single carbon atom vacancies in 3D stacked layered graphite via density functional theory (DFT) calculations. Palacios et al[6] calculated the electronic and magnetic structures of grapheme with mean-field Hubbard model taking single vacancies and voids, and showed the possibility of ferromagnetic as well as antiferromagnetic interactions among the induced moments. Yazyev et al[2] have studied the correlations among the magnetic moments at zigzag edges of graphene employing first principle calculations. Ordering of magnetic moments induced by extended line defects has also been explored by Ren et al[7].Experimental studies on the magnetism in graphene and graphitic systems has been pursued widely. Sepioni et al[8] and Nair et al[9] have established the presence of induced moments due to the defects and vacancies in graphene showing paramagnetism at low temperatures. The possibility of the presence of both ferro and antiferromagnetic orders in graphene samples has been well explored by Matte et al[10]. The presence of ferromagnetic order at room temperature[11] and its correlation with defects[12] in a graphite system has also been studied experimentally. The transport property of turbostratic graphene, a unique disordered stacking of tiny graphene-like units, is also well investigated [13,14].

Hybridization of graphene with h-BN layers, nitrogen or h-BN doped graphene, and the novel 2D material BCN have drawn significant attention of researchers lately in terms of their tunable electronic and magnetic properties. Several theoretical studies have been performed on the hybridization and doping scenario of h-BN and N in graphene or graphite system predicting the possibility of ferromagnetism and band gap tunability[15–22]. Novelty of the electronic structure and tunability of the band gap in the BCN system is nicely brought out by Kumar et al[23]. Novel bulk synthesis and the unique nature of the physical properties of the interesting BCN compound have been brought out in a series of combined important experimental and theoretical works by C. N. R. Rao and co-workers[24–29]. In the context of thin films, which is the most important form in the device context, Ci et al.[30] were the first to successfully grow the atomic layers of hybridized h-BN and graphene domains with the chemical vapor deposition (CVD) method. CVD is also widely used for growth of nitrogeneted graphene[31], BCN[32] and graphene/h-BN heterostructure[33]. Nitrogen ion assisted pulsed laser deposition (PLD)[34] and sequential ablation from h-BN and graphite targets with pulsed laser have also been reported for growing BCN films[35]. The electrical properties and the electronic structure of hybridized atomic graphene h-BN domains have been thoroughly studied by P. M. Ajayan and coworkers. A metal-insulator transition[36] and crossover to 2D Mott variable range hopping (M-VRH) as reflected by the deviation from Arrhenius activated behavior[37] in the low temperature regime have been reported by Ajayan group. The ferromagnetism triggered by nitrogen doping in graphene at room temperature is also well reported experimentally[38–40]. Zhao et al[41] reported the ferromagnetic character of carbon doped boron nitride above room temperature. In spite of the several important and interesting reports in the literature on the ferromagnetic nature and transport studies on these compounds the clarity regarding the inherent physics is still in debate, especially due to the diversity of sample forms and attendant issues pertaining to fruitful comparisons between them. The observed anomalous transition in temperature dependent magnetization has also not been clearly explained. In this work we attempt to connect and synchronize details of the electronic and magnetic phenomena in a concrete way in a pulsed laser deposited thin film platform for the BCN type material employing several spectroscopic techniques, transport and magnetization studies as well as first principle calculations.

## II. EXPERIMENTAL AND COMPUTATIONAL DETAILS

BCN films of thickness 500 nm are deposited on r-cut $Al_2O_3$ single crystal substrates with pulsed laser deposition (PLD) method using a KrF excimer laser (λ= 248 nm) with frequency 10 Hz and an energy density 2.5 J/$cm^2$. Target pellets are made by mixing of h-BN and graphite powders (Sigma- Aldrich USA) in different atomic ratios and pressing them at high pressure. For example, BCN-90 means the target is composed of 90 at.% carbon and 10 at.% h-BN. Three different targets are made with carbon at.% of 90, 98 and 100; the cases named as BCN-90, 98 and 100, respectively. All depositions are done at the substrate temperature 750°C and in presence of ammonia-argon gas mixture (30% ammonia) with chamber pressure $10^{-1}$ mbar. Pure carbon film is deposited in vacuum with base pressure $10^{-6}$ mbar. All films are annealed for 15 minutes at the growth temperature after deposition and cooled naturally. Thickness is measured with VEECO dektak 150 surface profiler.

Room temperature Raman Spectra are recorded on films using an upright confocal LabRAM HR 800 Raman Microscope (HORIBA Jobin Yvon SAS, rue de Lille, France) with 488 nm laser excitation. X-ray photoemission spectroscopy (XPS) measurements were performed using Al-Kα (E = 1486.7eV) lab-source. X-ray absorption near-edge structure (XANES) spectra at C, B and N K-edges were collected in total electron yield (TEY) mode. Room temperature valence band spectrum (VBS) was recorded with photon energy hν = 40 eV. Both XANES and VBS are performed using synchrotron radiation source at RRCAT, Indore at 'integrated photoemission spectroscopy beamline' (BL-02) in INDUS-1 and 'polarized light soft X-ray absorption spectroscopy beamline' (BL-01) in INDUS-2, respectively. Before recording the XPS, VBS and XANES spectra, the sample surface is sputtered using high energy Argon ions inside the chamber where base pressure is maintained at ~1 x $10^{-10}$ mbar in order to expose the uncontaminated surface. Resistivity measurements are performed by conventional four-probe method in the temperature range of 5 K- 300 K. Magnetization with respect to temperature and applied magnetic field is measured using commercial 7T-SQUID-VSM (Quantum Design Inc., USA) system. These measurements were performed with applied magnetic field parallel or in-plane to the film surface (IP). Magnetization vs. temperature was measured following the conventional protocols of zero field cooled warming (ZFC) and field cooled warming (FCW) cycles in the presence of applied magnetic field $\mu_0 H$ = 100 Oe.

The spin-polarized density functional theory (DFT) calculations[42,43] were performed within the projector augmented wave formalism,[44] for a plane-wave basis with a kinetic energy cut-off of 600 eV. The exchange-correlation energy was computed using the Perdew-Burke-Ernzerhof (PBE) form of generalized gradient approximation (GGA). [45]The lattice and ions were relaxed until the Hellman-Feynmann forces met the relaxation criteria of 0.015 eV/Å. The Brillouin zone was sampled using a Γ-centered 8×6×1 k-grid, and a dense mesh of Γ-centered 24×18×1 k-grid was used for the evaluation of density of states (DOS). Nitrogen and boron atoms were doped in a rectangular supercell of graphene containing 48 carbon atoms.

The periodic images of the supercell along the direction of vacuum were separated by 12 Å. The self-interaction error inherent to GGA functional results in spurious electron delocalization and incorrect magnetic solutions for both defects and dopants in graphene. To remedy this, we calculated the electronic and magnetic properties using the strongly constrained and appropriately normed (SCAN) meta-GGA functional[46].

## III. RESULTS AND DISCUSSION

### A. X-ray photoelectron spectroscopy (XPS)

Here we present and discuss the results of x-ray photoelectron spectroscopy (XPS) performed on two BCN thin films of different stoichiometries along with a carbon film, all grown by the pulsed laser deposition (PLD) technique, in ammonia atmosphere. The deposition of pure carbon film in ammonia was performed to serve as a reference for the degree of incorporation of nitrogen from the ambient during deposition as well as the specific role of boron on the film growth and stoichiometry. First we discuss the stoichiometry analysis based on the total areas of the fitted components, presented in the TABLE I.

TABLE I. Stoichiometry analysis from XPS fitting

| Sample | PLD target content | B1s area | C1s area | N1s area | N:C | B:C |
|---|---|---|---|---|---|---|
| BCN-90 | Target: 90 at.% graphite plus 10% BN | 747 | 9069 | 3114 | 0.34 | 0.08 |
| BCN-98 | Target: 98 at.% graphite plus 2% BN | 283 | 11760 | 1807 | 0.15 | 0.02 |
| BCN-100 | Target: 100 at.% graphite | ------ | 12557 | 10307 | 0.12 | ------- |

From the elemental analysis, the N:C ratio in BCN-100 (No BN in the PLD target) is 0.12 which implies that 12 at.% nitrogen gets got incorporated or fixed into the film from the gas phase ammonia during deposition. For BCN-98 (which has 2 at.% BN in the target) the ratio is a bit higher, as expected, due to the nitrogen contributed by the target in addition to that entering from the gas phase. In the BCN-90 case we can see significant incorporation of nitrogen (34 at.% ). This is more than the algebraic addition of 12 at.% entering from the gas phase and 10 at.% contributed by the target. This implies that boron helps in fixing nitrogen into the growing film.

Now we turn to the analysis of chemical states/bonding. In Fig. 1 for the BCN-100 case the C1s spectrum can be deconvoluted into two peaks. The peaks at 284.7eV and 286.3eV correspond to $sp^2$ C-C and C-N bonds, respectively. These peaks are very similar to those reported in Carbon Nitride films[47]. The N1s spectrum was fitted to a symmetric Gaussian form with the centre at 399.5eV implying significant contribution of C-N bonds in the sample, as expected.

Fig. 2 (a) and (b) show XPS core level spectra of C1s, B1s and N1s for the BCN-90 and BCN-98 films, respectively. In both the B1s spectra the feature is composed of two deconvoluted peaks. For the BCN-90 case the peaks are at 190.4 eV and 191.9 eV. The peak at lower energy corresponds to B-C bonds, while the peak at higher energy corresponds B-N bonds; as the electronegativity difference between B and C is lower than B and N[30]. For the BCN-98 case the peak for the B-C bond is shifted to 189.6 eV and the peak for the B-N bond is shifted to 192.4 eV. As the carbon concentration is more in this case, boron is likely to be bonded with carbon rather than nitrogen. The electronegativity of carbon is lower than nitrogen and hence the B-C bond energy gets shifted to lower side. For the same reason the bond energy for B-N is shifted to higher value as carbon is more electronegative than boron and it perturbs the B-N bond formation. Also, from the peak area it is clear that for the BCN-90 case the B-N bond formation is more favoured than the B-C bond formation, while in the BCN-98 case they are equally favourable. The whole feature is entirely different from the reported B1s spectra of h-BN[48] and hence one can conclude that there is no h-BN domain formation in both the cases.

The C1s spectra can also be deconvoluted into 3 peaks. For the BCN-90 case these peaks are located at 284.6 eV, 285.6 eV and 288.5 eV, which correspond to C-C, C-N, and C-B bonds, respectively.[49] The peak for the C-C bond is close to the peak observed for $sp^2$ C-C bond in graphite at 284.9 eV. For the BCN-98 sample the peaks are slightly shifted towards lower energy due to the presence of less nitrogen which is highly electronegative. From the peak area analysis it is clear that the main contribution is from the C-C bond due to the presence of large amount of carbon. Next is the contribution from C-N bond and for both the cases small amount of Boron contributes to the C-B bond density, which is less in BCN-90 sample than in BCN-98.

The N1s spectra are also composed of two deconvoluted peaks. For BCN-90 the peaks are at 399.3 eV and 401.1 eV. The peak at lower energy corresponds to the C-N bond and the peak at higher energy corresponds to the B-N bond. In the BCN-98 case the peak positions are almost the same as the chemical shift of 0.1 eV. Due to the presence of significant amounts of carbon and nitrogen in the system the C-N peak is prominent in both the cases and very close to the C-N peak observed in carbon nitride films at 399.8 eV in the form of sp2 coordinated pyridine like nitrogen[47]. As the presence of Boron is more in BCN-90, the B-N peak is also prominent in comparison with the BCN-98 case.

From the peak area analysis of the XPS data, it can be concluded that in both the BCN-90 and BCN-98 samples the amount of the bond formation is in order of C-C > C-N > B-N > B-C and that is reasonable as predicted from the theoretical model of BCN[23]. For the BCN-98 case B-C is more favoured than B-N as listed in the following TABLE II.

**TABLE II.** Bond analysis from XPS fitting

|         | C-C Bond Area | C-B Bond Area | C-N Bond Area | B-N Bond Area | C-N:C-C ratio | C-B: C-C ratio | B-N: C-C ratio |
|---------|---------------|---------------|---------------|---------------|---------------|----------------|----------------|
| BCN-90  | 4512.8        | 386.1         | 3168.5        | 653.9         | 0.70          | 0.09           | 0.14           |
| BCN-98  | 5895.3        | 940.6         | 2711.3        | 325.6         | 0.46          | 0.16           | 0.06           |
| BCN-100 | 7465.7        | -----         | 3266.2        | -----         | 0.44          | -----          | -----          |

Since the bonding nature of nitrogen is not fully and clearly understood from the XPS results, we resorted to the x-ray absorption near edge spectroscopy (XANES) measurements which give more detailed information about the local environment, as discussed in the next sub-section.

## B. X-ray absorption near edge spectroscopy (XANES)

Now we present and discuss the x-ray absorption near edge spectra (XANES) for the K edges of boron, nitrogen and carbon for the three samples of interest (Fig. 3 and Fig. 4). Many theoretical and computational approaches using Dipole Approximation, Quasi Particle Model and Multiple Scattering Theory are able to explain the experimental features successfully to certain extent[50]. Due to physical complexity in the real experimental scenario comparative study or analysis of dissimilarity between the shape, area and intensities of different K edges are strongly employed to study the local bonding environment[51].

The boron K edge (Fig. 3 (a)) is clearly more prominent in BCN-90 than in BCN-98 as the Boron content is very low in BCN-98. The peaks at 194.4 eV and 195.2 eV correspond to the 1s to $\pi^*$ transition due to B-N bond formation. The splitting is proposed to be due to the presence of triazinic and pyridinic nitrogen, respectively, which take part in the strong sp2 bond formation with Boron. The peak at 196.6 eV is due to the 1s to $\pi^*$ transition due to C-B bond formation.The peak is well shifted to higher energy in comparison to the B-C bond in the Boron Carbide compound. As with increasing C to B ratio in boron carbide compounds the C-B peak shifts to higher energy[52], here also the shift is more due to the presence of nitrogen with higher electronegativity. The absence of a strong signature at 192 eV [8,9] in both the cases implies that no domains of h-BN are present in the material to any discernible degree.The broad feature in the range 200 eV-210 eV corresponds to 1s to $\sigma^*$ transition. Here also we can observe the splitting of the peak due to B-N and B-C bonds, especially in the BCN-90 case.

So far as the carbon K edge (Fig. 4) for the three samples of interest is concerned, the peak at 288.8 eV corresponds to the 1s to $\pi^*$ transition. This peak is well shifted to the higher energy from the position observed in graphite, boron carbide[52] and in graphitic carbon nitride, which is around 285 eV[55,56]. This asymmetric peak is more close to the $\pi^*$ peak due to pyridinic[51] and triazinic nitrogen contributions, as observed in the $CN_x$ compounds[57]. The $\sigma^*$ feature can be distinguished into two different regions $\sigma_1^*$ and $\sigma_2^*$ in the ranges 290 eV to 293 eV and 295 eV to 301 eV, respectively. The $\sigma_1^*$ feature is composed of C-B or C-C and pyridinic C-N bond with maxima at 291 eV and 291.8 eV, respectively. As the C-C and C-B bond energies are very close, we cannot distinguish it between BCN-90 and BCN-98. In BCN-100 we can observe the background of C-C signature due to the absence of boron. The pyridinic C-N contribution is seen to be increasing with increasing nitrogen concentration from BCN-100 to BCN-90, as observed from the XPS data. The crucial role of boron in BCN compounds can be observed from the $\sigma_2^*$ feature. Here we can see the significant enhancement of triazinic C-N bond in BCN-98. Thus an optimum doping of 2 at.% boron in the framework triggers or favours the triazinic nitrogen which is lowest in the case of 10 at.% doping in BCN-90.

Calculating the $\sigma^*$ to $\pi^*$ intensity ratio as in TABLE III we can conclude that sp2 hybridization decreases with lowering the boron concentration in the material. For the special case of BCN-98 there is an optimum proportion ratio of the triazinic and pyridinic sp2 C-N bond which leads to the distinguishable magnetic and transport properties, as will be discussed later.

TABLE III. XANES peak analysis

| Carbon K edge | $I(\sigma_1^*)$ | $I(\sigma_2^*)$ | $I(\sigma_1^*): I(\pi^*)$ | $I(\sigma_2^*): I(\pi^*)$ |
|---|---|---|---|---|
| BCN-90 | 0.75 | 1.03 | 1.25 | 1.77 |
| BCN-98 | 0.65 | 1.10 | 0.79 | 1.34 |
| BCN-100 | 0.55 | 1.06 | 0.56 | 1.11 |

In the nitrogen K edge (Fig. 3 (b)) the feature at 407 eV is assigned to the $\pi^*$ transition which is close to pyridine[51] and triazine. The feature in the range of 409 eV to 420 eV corresponds to $\sigma^*$ transition[57]. Here also we can observe an enhancement of intensity and shift of maxima to higher energy in BCN-90 due to the pyridinic nitrogen contribution[51] as well as sufficient density of B-N bonds, as confirmed from the carbon K edge and XPS.

**C. Valence band spectroscopy (VBS)**

In Fig. 5 we present the Valence Band Spectra (VBS) of the three concerned BCN samples. The features seen are quite different as compared to those reported previously for graphene[58], carbon nitride[59], BN[60] and BCN films[61]. Here we can distinguish four different regions in the spectra. Region 1 between 1eV and 5eV represents the 2p π states, while region 2 between 5eV and 13eV represents the s-p hybridized states among carbon, boron and nitrogen. Region 3 from 13eV to 22eV denotes the mixed 2p σ and π states. Finally, the intense peak from 25eV to 32eV, region 4, emanates from the 2s σ states of the elements. Comparing the VBS data of the three samples of interest, we can observe a significant enhancement of region 2 corresponding to s-p hybridized states in the case of BCN-90 films as compared to the other two samples which is in conformity with the XANES data. In the BCN-98 case, the peak in region 2 is a bit sharper and shifted to lower energy and this could be due to the enhancement of the hybridization of triazinic nitrogen.The peak in the region 4 corresponding to the 2s σ states becomes sharper and shifted to higher energy with lowering boron as well as nitrogen concentration. As sp2 hybridization lowers the bonding energy in comparison to pure s-state, the feature suggests better growth of hybridized framework with increasing boron and nitrogen concentration. The broadening of the peak could be explained in terms of the mixing of $sp^2$ to pure s-states. In the BCN-100 case (no boron) the small feature at 34 eV which gradually becomes broadened and shifted to lower energy for BCN-98 and BCN-90 is not clearly understood. We used Raman spectroscopy in an attempt to elucidate these hybridization features and disorders in the material with greater clarity, as discussed in next sub-section.

**D. Raman Spectroscopy**

In Fig. 6 (a), (b) and (c) we show the room temperature Raman spectra of the three samples, BCN-100, BCN-98 and BCN-90, respectively, which are somewhat different as compared to previous report on bulk forms[26].The sharp peak around 1602 $cm^{-1}$ is denoted as G band which occurs due to the in plane optical phonon vibration $E_{2g}$ mode near the center of the Brillouin zone[62]. The strong signature of this mode implies the presence of sp2 hybridized 2 dimensional layers in the system[63]. The peak is significantly shifted to higher energy from the G band observed in graphite around 1580 $cm^{-1}$. Incorporation of strongly bonded nitrogen in the lattice plane results in a significant amount of bond stretching which affects the in-plane phonon vibration with the resultant Raman mode shifting to higher energy. This effect is very similar to the reported case of nitrogen doped grapheme sheets[64]. The asymmetric nature of the G band is predicted to emanates from the $D'$ mode arising due to the distortion that causes double resonance process in phonon scattering. The intense peak around 1360 $cm^{-1}$ is denoted as D band which occurs due to the $A_{1g}$ mode also resulting from the double resonance phonon scattering[63]. This mode is triggered by the induced disorder, defects and sp$^2$ bond breakage due to finite crystallinity and lattice distortion. Presence of this D band indicates the turbostraticity in the layered material. The band in the range 2500 $cm^{-1}$ to 3500 $cm^{-1}$ is referred to as 2D band which is basically an overtone of D band. As the Raman mode corresponding to the 2D band is extremely sensitive to in-plane stretching, strain and applied stress[65] we can observe the crucial role played by boron in the nitrogenation of the lattice from this mode. For all the three samples of interest the 2D band can be deconvoluted into three Gaussian contributions denoted as $2D_1$, $2D_2$ and $2D_3$. In TABLE IV the intensity ratios of D and 2D bands with respect to the G band are listed.

**TABLE IV**. Raman band analysis

|         | $I_D/I_G$ | $I_{2D_1}/I_G$ | $I_{2D_2}/I_G$ | $I_{2D_3}/I_G$ |
|---------|-----------|----------------|----------------|----------------|
| BCN-90  | 1.09      | 0.12           | 0.10           | 0.04           |
| BCN-98  | 1.24      | 0.11           | 0.11           | 0.03           |
| BCN-100 | 1.13      | 0.12           | 0.07           | 0.03           |

The disorder in turbostratic graphite like system can be mathematically related with D to G band intensity ratio through following equation.[63]

$$\frac{I_D}{I_G} = \frac{C(\gamma)}{L_a} \quad (1)$$

Here c(γ) is a constant for particular excitation energy and $L_a$ is the average distance among defect sites or the crystalline patches separated by grain boundaries. Taking c(γ) = 4.4 nm for excitation wavelength 488 nm we can calculate from the table that the average inter-defect distances are 4.04 nm, 3.55 nm and 3.89 nm for BCN-90 , BCN-98 and BCN-100, respectively. Thus, for the BCN-98 case the defect sites are denser than the other two cases resulting in the interesting electrical and magnetic features which are discussed in the next sub-sections. The splitting and broadening of the 2D band depends on several parameters like number of layers, inter layer interaction, layer orientation, excitation energy and induced strain in the plane[57]. In our case, as the excitation energy and thickness are the same for all the samples, the disorder and induced strain are the main responsible mechanisms here. From Figure 6 we can observe a sharp and intense splitting in the 2D band for BCN-98, suggesting that optimal 2 at.% doping of boron is inducing more strain in the nitrogeneted framework. The equality of $2D_1$ and $2D_2$ intensity ratio to G band could be due to the strain uniformity induced by pyridinic and triazinic nitrogens in the $sp^2$ bonded layer.

The overall abundance of pyridine and triazine-like configurations in addition to vacancy defects is clear from the present experimental results that are discussed already. Using the first-principles DFT approach, we therefore investigated the structural, electronic, and magnetic features of these individual defect configurations and their concurrent interactions.

E. First-principle DFT Calculations

The BCN thin films are modeled by doping boron and nitrogen atoms, as well as by introducing vacancy defects in a single layer graphene sheet. We emphasize on the nitrogen-doped configurations since the role of boron in modulation of overall electronic and magnetic properties of the films is secondary. Given the multiple pyridinic configurations, we focus on the trimerized pyridine as they are prevalent over mono and dimer type pyridines. [67]Among the vacancy defects, monovacancy configurations are considered due to their abundance and ability to induce a magnetic moment.[68]We examine the structural and electronic properties of monovacancy for a defect concentration of 2 at.%. [Fig. 7(a)] The removal of carbon atom leads to Jahn-Teller reconstructed 5–9 ring arrangement of the surrounding lattice where one carbon atom remains under-coordinated[69,70]. The electron deficient monovacancy gives rise to p-type behavior. An unpaired electron from the sp$^2$ state of the under-coordinated C atom, as well as the delocalized p$_z$ states from neighboring C atoms both contribute to the total magnetic moment of 2 μ$_B$ using the SCAN meta-GGA exchange-correlation functional, and the corresponding magnetization density is shown in Fig. 7(a). Such dual origin of vacancy magnetism is in agreement with the earlier experimental reports[68,71,72].

In the case of trimerized pyridine, N atoms replace the three C atoms surrounding the vacancy such that every N atom has two neighboring C atoms. Here [Fig. 7(b)], the concentration of N atoms is 6.38 at.%. Unlike the monovacancy defect, the trimerized pyridine configuration undergoes a symmetric relaxation due to an identical environment of lone pairs from N atoms surrounding the vacant site. The loss of p$_z$ electron associated with the missing atom results in p-type behavior. Further, compared to the pristine vacancy, the magnetic moment of trimerized pyridine is reduced to 1 μ$_B$ originating solely from the p$_z$ states.

Further, nitrogen with triazine configuration is another abundant defect complex, where three alternate C atoms in a single hexagonal ring are substituted by N [Fig. 7(c)]. As a result, the interaction between C and N atoms is in contrast to trimerized pyridine. In addition to the covalent C-N bond, a surplus electron density appears in the hexagonal ring, which is due to the charge transfer from the p$_z$ orbital of three C atoms within the triazine. A net imbalance of p$_z$ electrons between the two sub-lattices of graphene results in a magnetic moment of 1 μ$_B$ [Fig. 7(c)]. We report a half-metallic solution for the triazine configuration with N concentration of 6.25 at.%. This is in agreement with the prediction of half-metallicity in N/B-doped graphene[73,74]. Such spin-polarized carriers are promising for mediating an indirect magnetic exchange.

We have also considered B doped triazine-like and trimerized pyridine-like configurations. Both configurations have a magnetic moment of 1 $\mu_B$, similar to its N doped counterpart. However, the p-type metallic behavior observed in both B doped cases is in contrast to N doped triazine, which shows n-type behavior. Thus, we argue that in a nitrogen rich environment, the presence of boron may modulate the density of states at the Fermi level and affect the transport as well as long-range magnetic properties of the films.

**F. Transport Studies:**

Now we discuss the temperature dependent transport properties of the charge carriers in the BCN samples mentioned above. For a regular and perfect crystalline semiconductor, with increased temperature more electrons are pumped to the conduction band from the valence band, and the dc-resistivity is decreased as $\exp(E_g/K_B T)$, where $E_g$ is the activation energy and $K_B$ is Boltzmann constant. The scenario is quite different in semi-crystalline, or disordered systems. In such cases, the density of states is extended through the band gap due to the localised or trapped states because of the presence of local deformations, disorder in lattice, and defects. Over the full temperature range the transport mechanism is now controlled by two different mechanisms. Below a certain characteristic temperature and under low electrical bias the mechanism is fully governed by the hopping of charge carriers among these localised states. Above this temperature transport will be the usual band transport. The specific characteristic temperature can vary with the induced disorder in the system or more specifically with the density of localised states near $E_F$. Incorporating the possibility of hopping of carriers to sites with minimum hoping energy, irrespective of nearest neighbour or not, two most successful theoretical models of variable range hopping (VRH) have been propounded and employed to explain the experimental transport data. The main assumption behind the model proposed by Mott[75] is the approximation of constant density of states around the Fermi level, $N(E_F)$ irrespective of energy of the system. Efros and Shklovskii on the other hand showed that below a certain critical temperature the density of states may not be constant and it becomes non uniform near $E_F$ due to the coulomb interactions among carriers[76]. Incorporation of this interaction causes a dip in the density of states around $E_F$ known as the Coulomb gap. Both model can be described mathematically by the following equation[77].

$$\rho = \rho_0 \exp(T_0/T)^\upsilon \qquad (2)$$

where $\rho$ is the dc-resistivity, $\rho_0$ is a constant, and $T_0$ is the characteristic temperature. In the Mott model the value of $\upsilon$ is 1/3 and 1/4 for 2D and 3D systems, respectively; while $\upsilon = 1/2$ for all the systems in the ES model which possesses the Coulomb gap.

In Fig. 8 we present the transport data for our BCN-90, BCN-98 and BCN-100 samples. Temperature dependent normalised dc resistivity data for the samples are shown in Fig. 8 (a) for comparison. For the determination of the hopping mechanism in each case, the curves are nonlinearly fitted directly to equation (2) with varying the parameters $\rho_0$, $T_0$ and $\upsilon$ as shown in Fig. 8 (b), (c) and (d). The important parameters are tabulated in TABLE V.

**TABLE V**. Nonlinear fitting parameters

|  | BCN-90 | BCN-98 | BCN-100 |
|---|---|---|---|
| $T_0$(K) | 1226.1 | 94.1 | 311.2 |
| $\upsilon$ | $0.536 \pm 0.001$ | $0.442 \pm 0.002$ | $0.260 \pm 0.004$ |

Here we can observe that the value of $\upsilon$ is very close to 1/2 for BCN-90 and it is close to 1/4 for BCN-100. It is clear that for BCN-90 the hopping of carriers is governed by the Mott model over the entire temperature range and the same can be understood from the fact that the characteristic temperature $T_0$ is 1226 K which is far above our measurement range. In the BCN-100 case the ES model fits the data very well over the whole range, as $T_0$ is above room temperature. The most interesting case is the BCN-98 sample (as was also discussed in the spectroscopic analysis) where υ does not match either with 1/2 or 1/4 and it lies between these values. This crossover between Mott and ES Scenarios is very clear from the value of the critical temperature $T_0$ of 94 K that lies within the concerned temperature range of our measurement. For further clarification, the low temperature data up to the crossover temperature was fitted (Fig. 7 (c) inset) to equation (2) and the fitted value of υ was found to be 0.506±0.002 which is closer to 1/2 consistent with the ES model. Below the critical temperature the Coulomb interaction becomes dominant resulting into the applicability of the ES process.[78].

From the whole picture the crucial role played by boron in the electronic structure is thus very important. By increasing the boron doping concentration in nitrogenated turbostratic carbon framework, the hopping mechanism of charge carriers among the localized states is seen to be changing dramatically with a clear crossover for an optimum 2 at.% doping as in BCN-98. From a continuous and constant density of states $N(E_F)$ near Fermi level with no boron doping, we can get a transition to a clear Coulomb gap state with increasing boron concentration in the material. The carrier localisation length ($\varsigma$) and the $N(E_F)$ can be related to the calculated average inter-defect sites distance ($L_a$) from Raman spectroscopy and the prediction from first principle calculations.

## G. Magnetic Measurements

Isothermal field dependent magnetization data for all the three BCN samples of interest are presented in Fig. 9 (a) and (b) at 300K and 5K, respectively. Diamagnetic contribution from the substrate was subtracted from the raw data by following the standard procedure of fitting to the large field linear contribution. We can observe from these data that at room temperature (300 K) ferromagnetism is present in all the cases with small coercive field values ($H_C$) and saturation magnetizations ($M_S$), and the two are anti-correlated. At 5K the $M_S$ increases in each case but magnetizing the sample is seen to get harder, as expected for the case of the magnetic system with disorder. Disorder can render spin canting in the system, which resists the saturation of the net magnetization, very similar to a spin glass system[79]. The disorder in the present case is related to the nitrogen dopant incorporation and the specific nature of disorder depends on the defect type and its consequence for vacancy defect creation due to issues of charge balance. It is thus of interest to discuss the nitrogen concentration dependence of the magnetization parameters.

In Fig. 9 (c) and (d) we present the $H_C$ and $M_S$ values with respect to the nitrogen concentration present in the system for all the three samples at 300K and 5K, respectively. At room temperature, the saturation magnetization increases while the coercive field decreases monotonically with the increase in the nitrogen concentration from BCN-100 to BCN-90. As from the spectroscopic analysis, since both the triazinic and pyridinic sp$^2$ C-N bonds lead to induced spins at the defect sites, increase of the nitrogen concentration can certainly lead to an higher $M_S$. If we consider this as an inhomogeneous ferromagnet with the presence of disorder, more in the form of correlated ferromagnetic clusters, increase in the spin density with increasing nitrogen concentration can render enhanced correlations and larger cluster size thereby decreasing the coercive field.

At 5K the behaviour of $M_S$ is the same as at 300K, albeit with increased saturation magnetization in each case, but interestingly can observe a huge increase in $H_C$ for BCN-98. This anomalous increase appears to be directly connected to the increase of triazinic sp$^2$ C-N bond concentration reflected in XANES spectra for this specific case. This ferromagnetic condition of the sample is reminiscent of single domain magnetic clusters, which give very high coercive field value. The low coercive field for the BCN-90 case, wherein the nitrogen defect concentration is substantially higher, reflects the case of a more homogeneous ferromagnetic system.

In Fig. 10 we present the data for temperature dependent magnetization. Fig. 10 (a), (b) and (c) represent the zero field cooling (ZFC) and field cool warming (FCW) data of M vs T for BCN-90, 98 and 100 samples, respectively, with an applied field 100 Oe. We can see that the bifurcation between the ZFC and FCW curves increases with the increasing nitrogen (and boron) contents. This indicates the nature of evolution of the magnetic state with increased spin density in the presence of disorder. The bifurcation temperature is almost near room temperature in all cases including the case of BCN-100 wherein there is no boron. The BCN-98 and BCN-100 samples both show ferromagnetic ordering developing near 130 K, which is distinctly defined in the BCN-98 case. In the BCN-90 case this transition is not clearly seen possibly because of its significant broadening due to enhanced dopant concentration and related spin disorder. In this case of BCN-90 a sharper downturn in ZFC and inflection in FCW is noted around 65K. At about the same temperature even in the BCN-98 sample an upturn is seen in FCW. We plotted the magnetization Vs resistivity for the BCN-90 and BCN-98 samples to explore if there is any connection of the upturn with the carriers in the system. The data are shown in Fig. SI-2. In both the cases an inflection point is noted around 65K, although it is sharper in the BCN-90 case. When we examined the magneto-transport in all the three systems (see Fig. SI-3) we did not find any significant magneto-resistance (MR) even in high field. This means that the transport connection may occur only via carrier density aspect instead of scattering aspect, and the same may change due to trapping (localization) of carriers below a certain temperature (e.g. 65K). The trapping of charge by a nitrogen defect may influence its spin state and thereby change the magnetization systematic below this temperature. Further studies will be clearly needed to elucidate these aspects in this rather intricately complex system.

Behaviour of FCW susceptibility ($\chi_{FCW}$) and inverse of $\chi_{FCW}$ of the samples are shown in Fig. 10 (d) and (e). As our materials are weakly ferromagnetic at room temperature and are not homogeneous ferromagnets, they do not follow the Curie-Weiss law in our temperature range. To understand the magnetic transition more clearly the derivative of $\chi_{FCW}$ with respect to temperature is plotted in Fig. 10 (f). For the special case BCN-98, as mentioned earlier we can see the drastic change of the derivative around 130 K resulting in a strong dip in the plot. This is indicative of the clear magnetic transition.

In order to gain some insights into the microscopic mechanism of the induced spin due to the presence of different type of C-N bonds and possible role for boron, we resorted to the density functional theory (DFT) calculations, as addressed below. Specifically, we attempt to elucidate the microscopic mechanism of long-range magnetism by considering interactions among the different defect complexes. Previous reports on magnetically ordered N-doped graphene have attributed the behaviour to an abundance of pyridinic and substitutional N complexes separately[38,80]. However, the synergistic effects of interacting defects on the overall magnetism have not been explicitly addressed, which are essential to understand the long-range magnetic ordering. We consider the interaction of trimerized pyridine with other defects since trimerized pyridine is present in significant quantities. A maximum N concentration of 15% is considered in the following discussion.

For two trimerized pyridines in close proximity, the spin polarization is mainly due to $p_z$ states and an additional fractional moment is induced on $sp^2$ states of N atoms, which contributes as a paramagnetic centre. [Fig. 11(a)] Investigating the density of states we find that the nitrogen $sp^2$ states appear as mid-gap states. With increasing the distance between the two pyridinic defect complexes, the $sp^2$ contribution to the magnetism vanishes and only $p_z$ states generate a semi-local moment mediated via direct exchange. For a pyridinic defect interacting with neighboring monovacancy, the $p_z$ spins align parallel (antiparallel) when the two defects are on the same (different) graphene sublattices. The unpaired $sp^2$ state of the vacancy retains its local moment and is robust to carrier doping and/or structural changes in the presence of neighboring defect. [Fig. 11(b)] In addition, we also report magnetic solutions for interacting monovacancy with mono and dimerized pyridine. Lastly, the magnetic interaction between the triazine and pyridine defect complexes generates both ferromagnetic and antiferromagnetic order depending on the sub-lattice site. [Fig. 11(c)] Interestingly, the p-doped pyridine alters the half-metallic behaviour of triazine, however an indirect exchange is still possible due to the presence of $p_z$ conduction electrons at the Fermi level.

Thus, we propose the primary mechanism driving the ferromagnetic order is a direct exchange among the nitrogen defect configurations due to the high concentration of N atoms in the turbostratic BCN thin films. The spatial extent of ~2 nm for the unpaired $p_z$ states of monovacancy exceeds the experimentally estimated inter-defect distances[3]. However, the sensitivity of $p_z$ states towards disorder can limit the role of direct exchange. Further, it is predicted that disorder in pure graphene leads to Anderson insulating behaviour for vacancy defect with higher concentration, and a substantial gap opening could originate in case of N substitution[81,82] In such a scenario, the scattering of conduction electrons will result in loss of carrier-mediated indirect exchange pathways. However, the formation of bound magnetic polarons in monovacancy and pyridinic defects is another possibility of inducing ferromagnetic order through an indirect exchange that does not depend on conduction electrons[83,84].

## IV. Conclusions

Thin films of $B_xC_yN_z$ compound are grown using the Pulsed Laser Deposition (PLD) method, and their magnetic and transport properties are measured and analyzed in their entirety using various spectroscopic techniques like X-ray photoelectron spectroscopy (XPS), Valence Band Spectroscopy (VBS) and X-ray absorption near edge spectroscopy (XANES). Theoretical insights are obtained using Density Functional Theory (DFT) calculations. A dramatic crossover in the transport mechanism of charge carriers is noted with the change in the doping level of specific nitrogen defects. A robust and high saturation magnetization is achieved in the $B_xC_yN_z$ films, which is seen to be higher by almost hundred times as compared to the similarly grown undoped turbostratic graphene film (Fig. SI-1). An anomalous transition and increase in the coercive field is also observed at low temperature. The detailed comparison of the experimental data with the theoretical results brings out the intriguing role of specific nitrogen defects in defining the physical properties of this interesting material.

## Acknowledgement

R.M. and S.O. acknowledge Dr. Rajeev Rawat from UGC-DAE CSR Indore for providing the facility for temperature dependent resistivity measurement and Dr. D. M. Phase from UGC-DAE CSR Indore for helping with the measurements in synchrotron beamline at RRCAT, Indore. M. K. acknowledges the funding from the Department of Science and Technology, Government of India under Ramanujan Fellowship. M.K. and S.O. acknowledge DST Nano Mission Thematic Unit project, SR/NM/TP-13/2016, for funding.

**Figure Captions:**

**Fig 1:** C1s and N1s core x-ray photoelectron spectra of BCN-100

**Fig 2:** C1s, B1s and N1s core x-ray photo electron spectra of (a) BCN-90 and (b) BCN-98

**Fig 3:** X-ray near edge spectra for (a) Boron K edge and (b) Nitrogen K edge of BCN-90, 98 and 100

**Fig 4**: X-ray near edge spectra for Carbon K edge of BCN-90, 98 and 100

**Fig 5:** Valance Band Spectra of BCN-90,98 and 100

**Fig 6:** Raman Spectra and peak fitting for 2D band splitting of (a) BCN-100, (b) BCN-98 and (c) BCN-90

**Fig 7:** (L-R) Atomic arrangement, spin charge density for the (a) monovacancy, (b) trimerized pyridine, and (c) triazine configurations. Black (light blue) spheres correspond to carbon (nitrogen) atoms. The positive (negative) spin charge densities are denoted by red (blue) color.

**Fig 8:** (a) Normalized resistivity vs temperature. (b), (c) and (d) represent the temperature dependent resistivity curve with non linear fitting. Linear fits are shown for (b) and (d) (inset). Non linear fit is shown for small range bellow the crossover for BCN-98 in (c) (inset).

**Fig 9:** (a) and (b) represent the isothermal M-H loop for three BCN Samples at 300 K and 5 K respectively. (c) and (d) represent the dependence of coercivity ($\boldsymbol{\mu_0 H_C}$) and saturation magnetization($\boldsymbol{M_S}$) on nitrogen concentration in films at 300 K and 5 K respectively.

**Fig 10:** (a), (b) and (c) represent the ZFC and FCW curves for BCN-90,98 and 100 respectively. (d) normalised susceptibility vs temperature. (e) inverse susceptibility vs temperature (f) derivative of susceptibility vs temperature.

**Fig 11:** Spin charge densities for the interacting pairs of (a) trimerized pyridine - trimerized pyridine, (b) trimerized pyridine - monovacancy and (c) trimerized pyridine - triazine configuration.

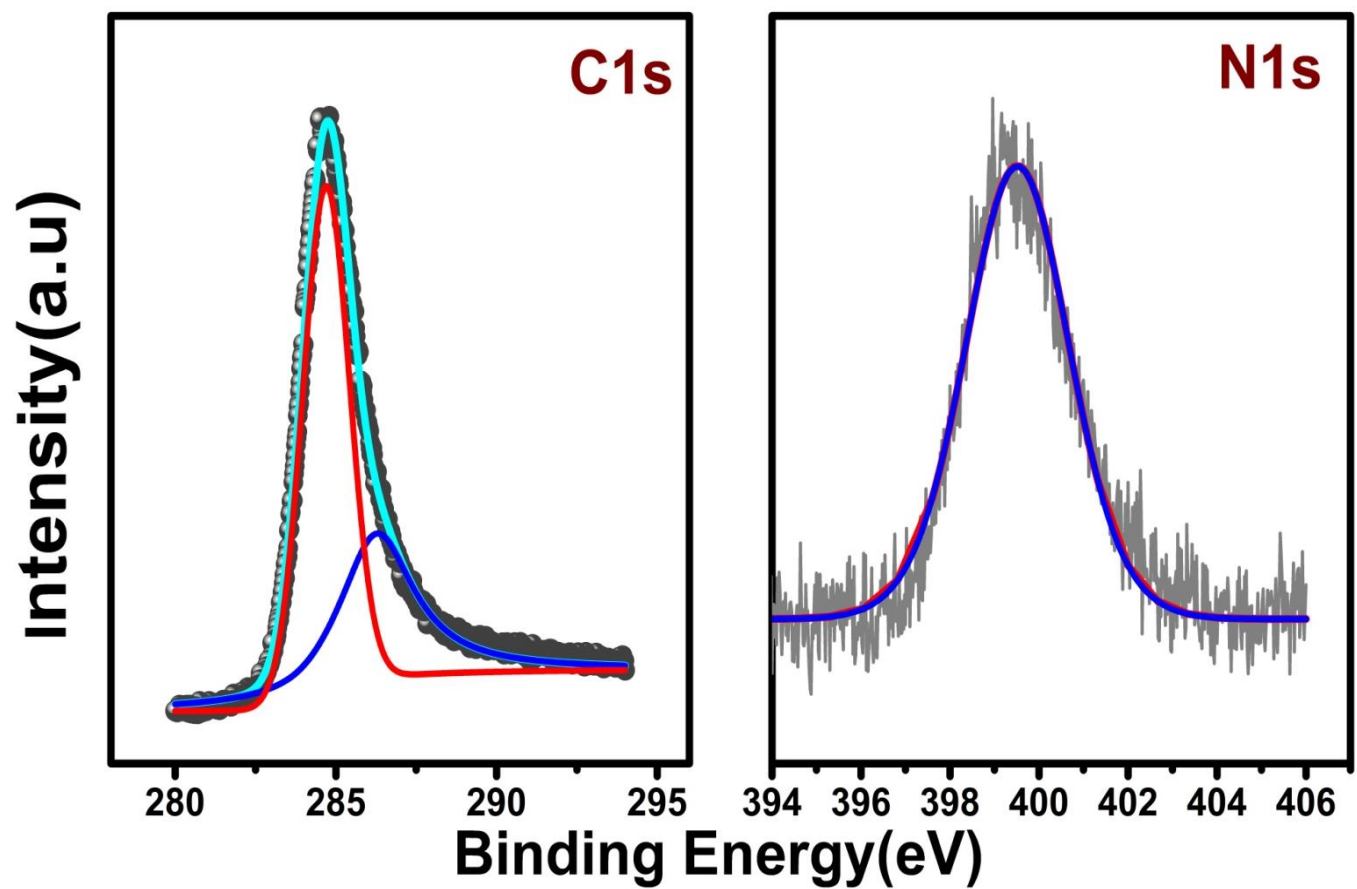

**Figure 1**

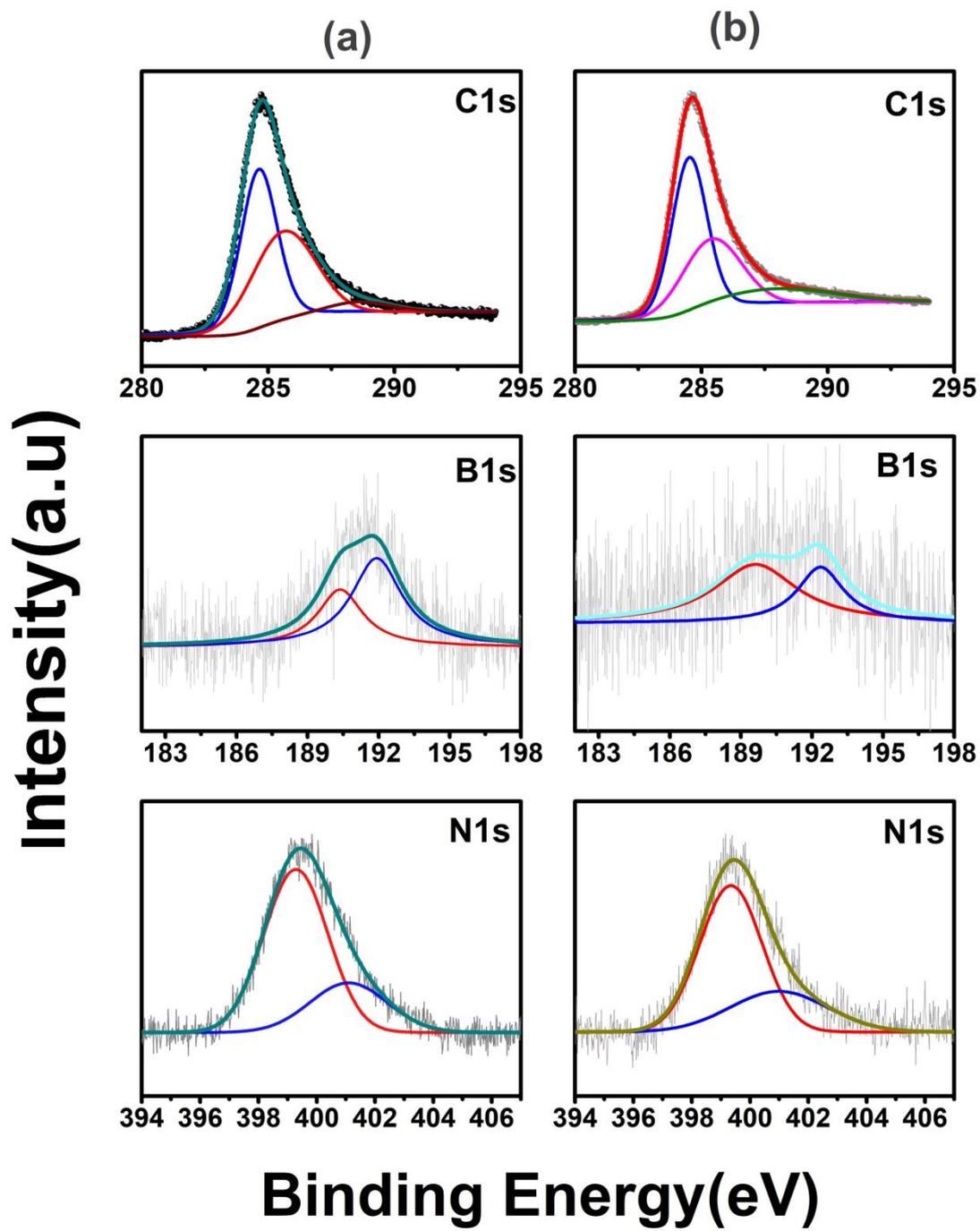

**Figure 2**

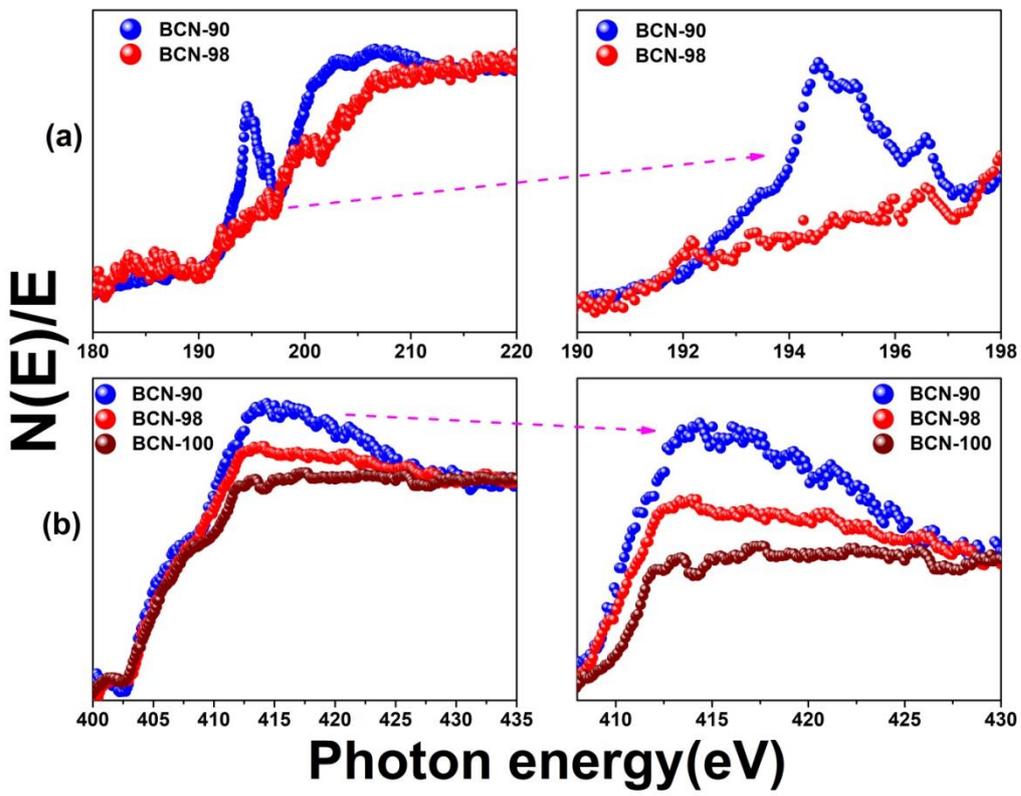

Figure 3

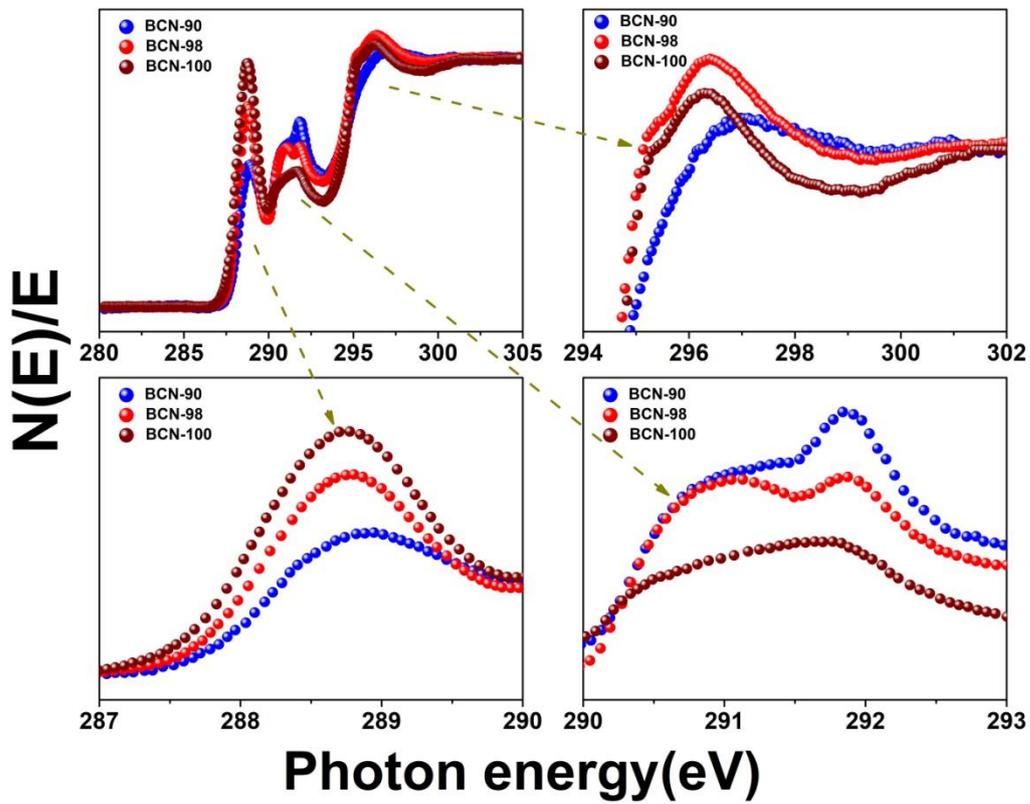

Figure 4

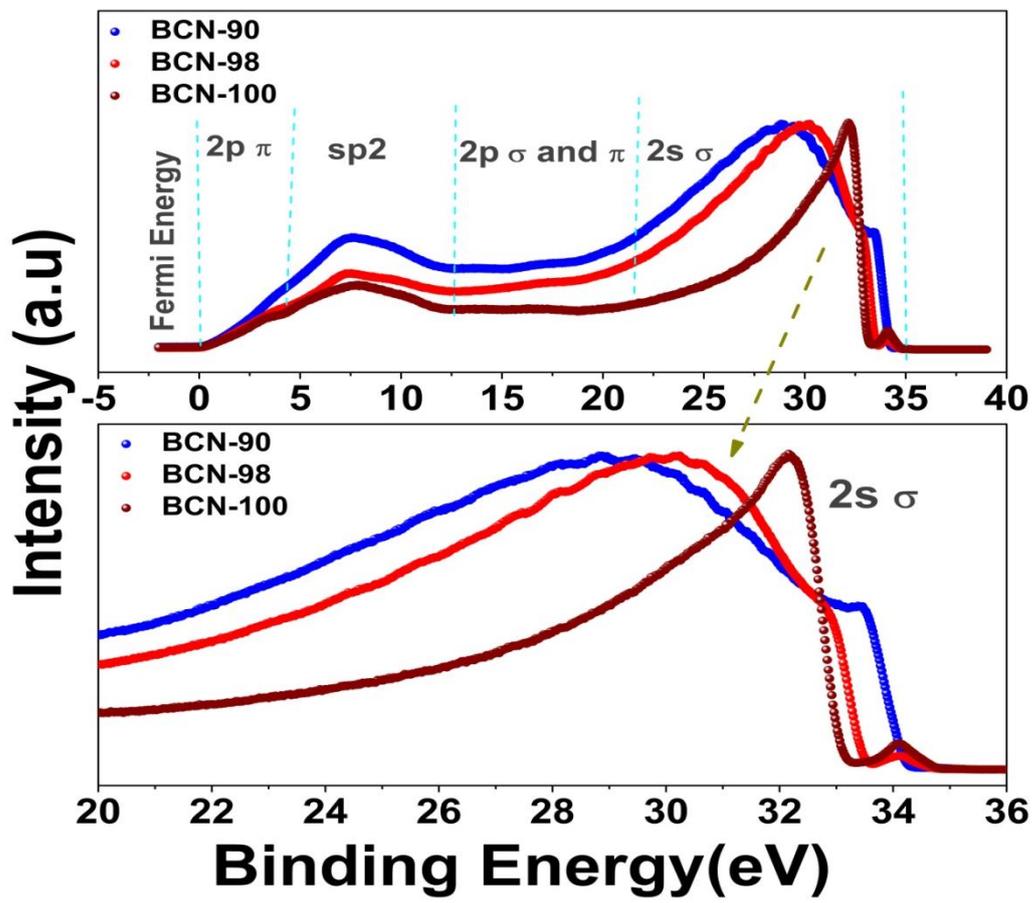

**Figure 5**

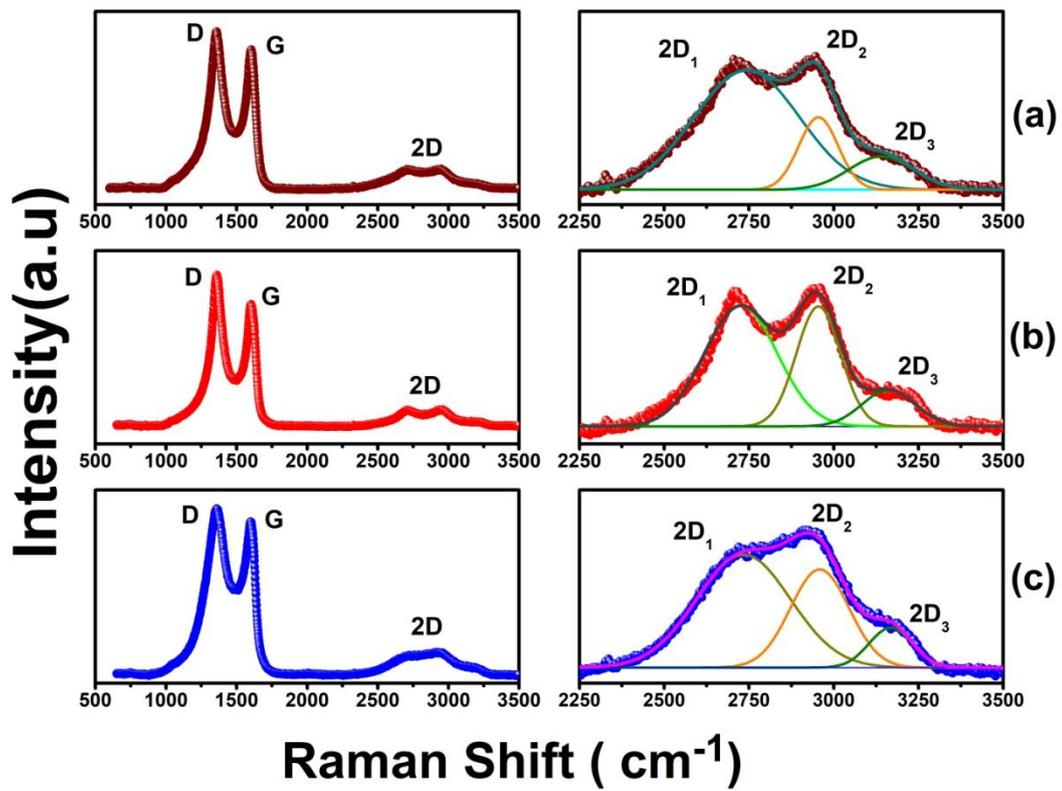

**Figure 6**

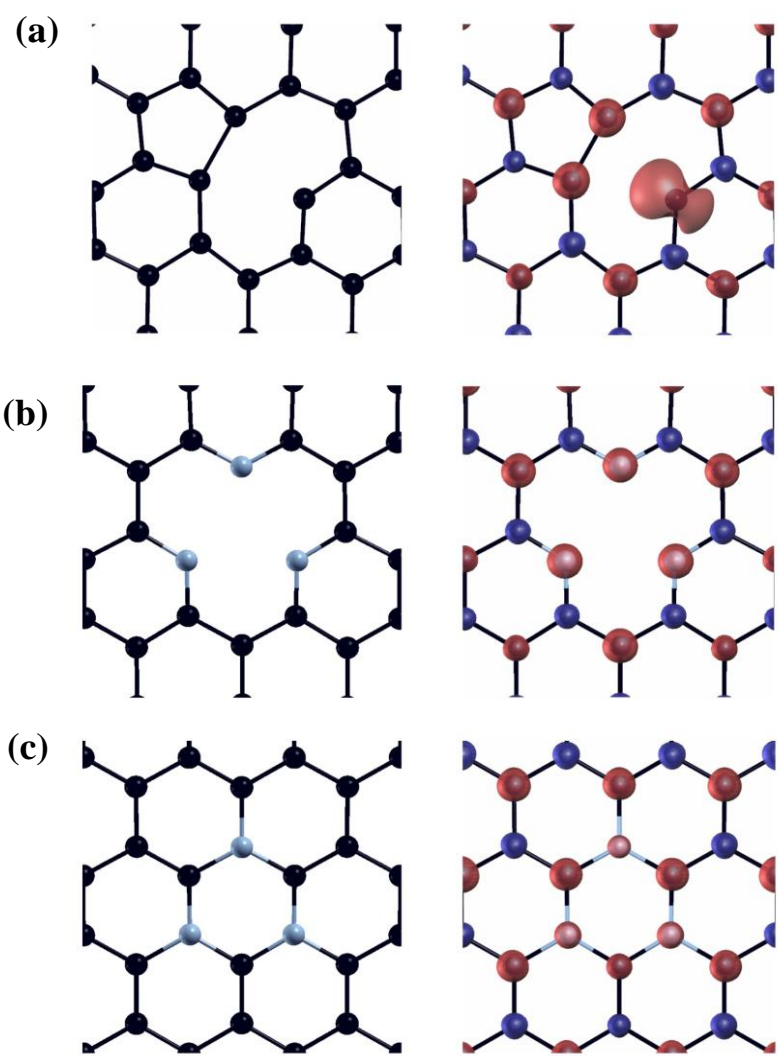

**Figure 7**

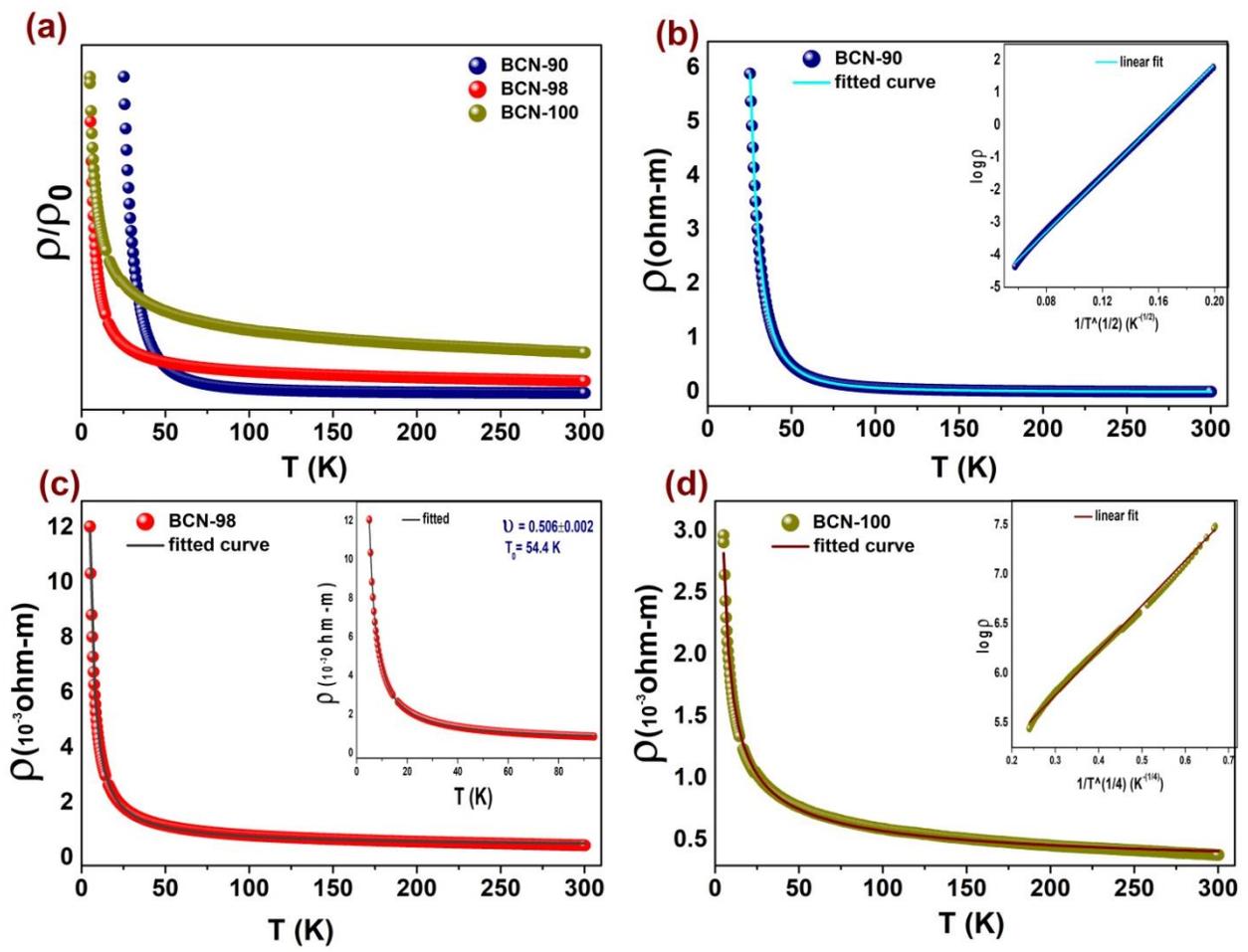

**Figure 8**

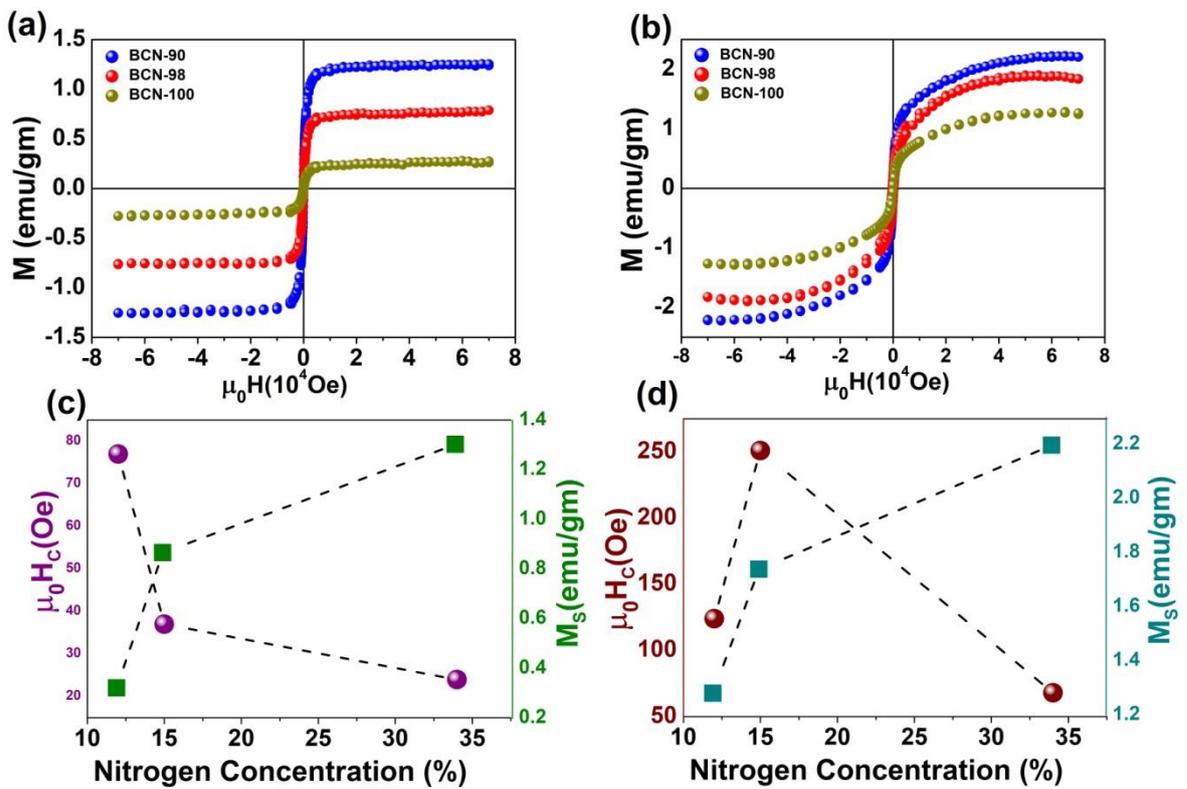

**Figure 9**

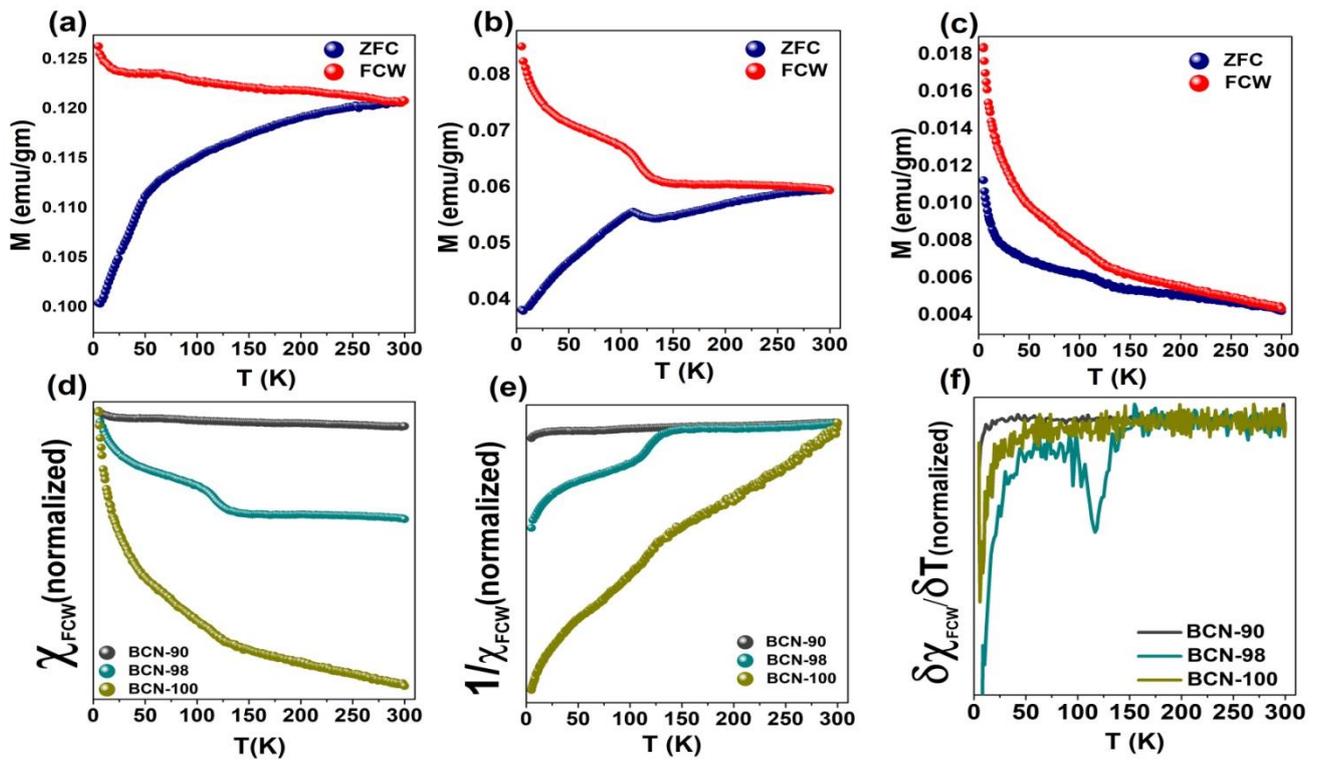

**Figure 10**

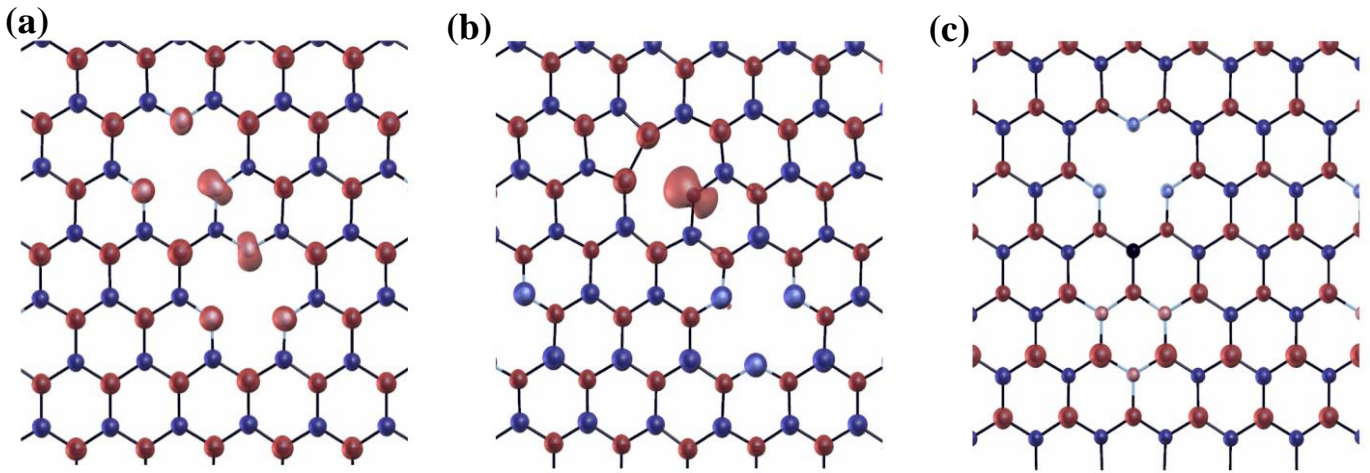

**Figure 11**